\documentclass[a4paper,12pt]{article}
\usepackage{amsmath, bm}
\usepackage{pstricks}
\usepackage{pst-node}
\usepackage[ansinew]{inputenc}
\usepackage{amssymb,amsmath}
\usepackage{amsfonts}
\usepackage{epsfig}

\usepackage{pst-node}

 \let\~\widetilde
\def\BSE{\begin{subequations}}\def\ESE{\end{subequations}}
\let \= \equiv

\def\p{\partial}
\def\px{\partial_x}
\def\py{\partial_y}
\def\a{\alpha}
\def\b{\beta}
\def\g{\gamma}
\def\d{\delta}
\def\o{\omega}
\def\wt{\widetilde}
\def\ms{\medskip}

\font\Sets=msbm10

\def\Integer {\hbox{\Sets Z}}    \def\Real {\hbox{\Sets R}}

\def\Complex {\hbox{\Sets C}}   \def\Natural {\hbox{\Sets N}}

\def\be{\begin{equation}}       \def\ba{\begin{array}}

\def\ee{\end{equation}}         \def\ea{\end{array}}

\def\bea {\begin{eqnarray}}      \def\eea {\end{eqnarray}}

\def\bean{\begin{eqnarray*}}    \def\eean{\end{eqnarray*}}

\def\pa  {\partial}             \def\ti  {\widetilde}

\def\la  {\lambda}

\def\eps{\varepsilon}           \def\ph{\varphi}

\def\const {\mathop{\rm const}\nolimits}

\def\tr    {\mathop{\rm trace}\nolimits}

\def\res   {\mathop{\rm res}  \limits}

\def\diag  {\mathop{\rm diag} \nolimits}

\def\im    {\mathop{\rm Im}   \nolimits}

\def\ker  {\mathop{\rm Ker} \nolimits}

\def\RA {\ \Rightarrow\ }         \def\LRA {\ \Leftrightarrow\ }

\def\qed   {\vrule height0.6em width0.3em depth0pt}

\def\<{\langle} \def\({\left(}  \def\>{\rangle} \def\){\right)}

\def\defeq {\stackrel{\mbox{\rm\small def}}{=}}

\author{Elena Kartashova$^*$, Alexey Kartashov$^\dag$ \\
\\
$*$ RISC, J.Kepler University, Linz, Austria\\
$^\dag$ AK-Soft, Linz, Austria\\
\\\\  e-mails:\\ lena@risc.uni-linz.ac.at, alexkart1@gmx.at
}

\title{LAMINATED WAVE TURBULENCE: GENERIC ALGORITHMS II}

\begin{document}

\maketitle

\begin{abstract}

The model of laminated wave turbulence puts forth a novel
computational problem - construction of fast algorithms for finding
exact solutions of Diophantine equations in integers of order
$10^{12}$ and more. The equations to be solved in integers are
resonant conditions for nonlinearly interacting waves and their form
is defined by the wave dispersion. It is established that for the
most common dispersion as an arbitrary function of a wave-vector
length two different generic algorithms are necessary: (1)
one-class-case algorithm for waves interacting through scales, and
(2) two-class-case algorithm for waves interacting through phases.
In our previous paper we described the one-class-case generic
algorithm and in our present paper we present the two-class-case
generic algorithm.

\end{abstract}

\maketitle

\def\p{\partial}
\def\px{\partial_x}
\def\py{\partial_y}
\def\a{\alpha}
\def\b{\beta}
\def\g{\gamma}
\def\d{\delta}
\def\o{\omega}
\def\wt{\widetilde}
\def\ms{\medskip}

\font\Sets=msbm10

\def\Integer {\hbox{\Sets Z}}    \def\Real {\hbox{\Sets R}}

\def\Complex {\hbox{\Sets C}}   \def\Natural {\hbox{\Sets N}}

\def\be{\begin{equation}}       \def\ba{\begin{array}}

\def\ee{\end{equation}}         \def\ea{\end{array}}

\def\bea {\begin{eqnarray}}      \def\eea {\end{eqnarray}}

\def\bean{\begin{eqnarray*}}    \def\eean{\end{eqnarray*}}

\def\pa  {\partial}             \def\ti  {\widetilde}

\def\la  {\lambda}

\def\eps{\varepsilon}           \def\ph{\varphi}

\def\const {\mathop{\rm const}\nolimits}

\def\tr    {\mathop{\rm trace}\nolimits}

\def\res   {\mathop{\rm res}  \limits}

\def\diag  {\mathop{\rm diag} \nolimits}

\def\im    {\mathop{\rm Im}   \nolimits}

\def\ker  {\mathop{\rm Ker} \nolimits}

\def\RA {\ \Rightarrow\ }         \def\LRA {\ \Leftrightarrow\ }

\def\qed   {\vrule height0.6em width0.3em depth0pt}

\def\<{\langle} \def\({\left(}  \def\>{\rangle} \def\){\right)}

\def\defeq {\stackrel{\mbox{\rm\small def}}{=}}

\newpage

\section{INTRODUCTION}

The most general problem setting of the wave turbulence theory can
be regarded in the form of a nonlinear partial differential equation
$$
\mathcal{L}(\psi)=\varepsilon \mathcal{N}(\psi)
$$
where $\mathcal{L}$ and $\mathcal{N}$  denote  linear and nonlinear
part of the equation correspondingly, the linear part has the standard
wave solutions of the form
$$
\varphi = A \exp {i[ \vec{k}\vec{x} - \o t]},
$$
where the amplitude $A$ may depend on space variables but not on
time,
 and a small parameter $\varepsilon$ shows that only
resonant wave interactions are taken into account. The dispersion
function $\o=\o(\vec{k})$ can be easily found by substitution of
$\varphi$ into the linear part of the initial PDE, $
\mathcal{L}(\varphi)=0,$ while $ \p_t \leftrightarrow i \o \quad
\mbox{and} \quad \p_{x_s} \leftrightarrow i k_s, $ and resonance
conditions have the form
\begin{eqnarray}\label{open}
\begin{cases}
\omega (\vec k_1) \pm \omega (\vec k_2)\pm ... \pm \omega (\vec k_{n}) = 0,\\
\vec k_1 \pm \vec k_2 \pm ... \pm \vec k_{n} = 0.
\end{cases}
\end{eqnarray}
for $n$ interacting waves with wave-vectors $\vec k_{i}, \ \
i=1,2,...,n$. For most physical applications it is enough to regard
$n=3$ or $n=4$, and the most common form of dispersion function is
$$ \o= \o(|\vec{k}|), \ \ \ \ |\vec{k}|=\sqrt{m^2+n^2} \quad \mbox{for} \quad \vec{k}=(m,n)$$
 (for instance, capillary, gravitational and surface
water waves, planetary waves in the ocean, drift waves in tokamak
plasma, etc.)\\

The model of laminated wave turbulence\cite{lam} describes two
co-existing layers of turbulence - continuous and discrete - which
are presented by real and integer solutions of Sys.(\ref{open})
correspondingly. The continuous layer is described by classical
statistical methods\cite{lvov} while for the discrete layer new
algorithms have to be developed. It was shown in\cite{AMS} that an
arbitrary integer lattice $(m,n)$, each node of the lattice denoting
a wave-vector $\vec{k}=(m,n)$, can be divided into some clusters
(classes) and there are two types of solutions of Sys.(\ref{open}):
those belonging to the same class and those belonging to different
classes. Mathematically, a class is described as a set of
wave-vectors for which the values of the dispersion function have the
same irrationality. For instance, if the dispersion function has the
form $\o=\sqrt{m^2+n^2}$, then a class is described as follows:
$$\{ m_i,n_i\}: \ \sqrt{m_i^2+n_i^2}=\g_i\sqrt{q}$$
where $\gamma$ is a natural number and
$q$ is a square-free integer. Physically, it means that waves
are interacting over the scales, that is, each two interacting waves
generate a wave with a wavelength different from the  wave lengths
of the two initial waves. Interactions between the waves of
different classes do not
generate new wavelengths but new phases.\\

In our preceding paper\cite{KarKar} we presented a generic algorithm for
computing all integer solutions of Sys.(\ref{open}) within one
class. Four-wave interactions among 2-dimensional gravitational
water waves were taken as the main example, in this case
Sys.(\ref{open}) takes form:
\bea\label{prosetdef2} \begin{cases}
{(m_1^2+n_1^2)}^{1/4} + {(m_2^2+n_2^2)}^{1/4}={(m_3^2+n_3^2)}^{1/4}+{(m_4^2+n_4^2)}^{1/4} \\
m_1+m_2=m_3+m_4\\
n_1+n_2=n_3+n_4\\
\end{cases}\eea
and classes are defined as $Cl_q=\{\g^4q\}$, where $q$, called class
index, are all natural numbers containing every prime factor in
degree smaller $4$ and $\g$, called weight, all natural numbers. It
can be proven that if all 4 wave-vectors constructing a solution of
Sys.(\ref{prosetdef2}) {\bf do not} belong to the same class, then
the only possible situation is following: all the vectors belong to
two different classes $Cl_{q_1},Cl_{q_2}$ and  the first equation of
Sys.(\ref{prosetdef2}) can be rewritten then as
\be\label{th2eq3} \g_1\sqrt[4]{q_1} +
\g_2\sqrt[4]{q_2}=\g_1\sqrt[4]{q_1}+\g_2\sqrt[4]{q_2} \ee
with some $\g_1, \g_2 \in \Natural$ and $q_1, q_2$ being class
indexes. In the present paper we deal with this two-class case.

\section{COMPUTATIONAL PRELIMINARIES}

As in the previous paper \cite{KarKar}, we are going to find all solutions of
Eq.(\ref{prosetdef2}) in some finite domain $D$, i.e. $|m_i|, |n_i|
\le D$ for some $D\in \Natural$. The first case has been studied for
$D=1000$, where $\pi_{cl}(10^3) = 384145$ classes have been
encountered. The straightforward approach, not making use of
classes, consumes, as for the first case, at least $O(D^5)$
operations and is out of question (see \cite{KarKar}, Sec.3.2.1 for
discussion of this
point).\\

Straightforward application of classes also does not bring much. The
Eq.(\ref{th2eq3}) is now trivial - but classes are interlocked
through linear conditions. Even if for each pair of classes we could
detect interlocking and find solutions, if any, in $O(1)$ operations
(which is probably the case, though we did not prove it), the
overall computational complexity is at least $\pi_{cl}(D)^2$ - i.e.
not much less than $O(D^4)$. For $D=1000$ this implies
$1.5\cdot10^{11}$ pairwise class matches which is
outside any reasonable computational complexity limits.\\

The trouble with this approach - as, for that matter, with virtually
any algorithm consuming much more computation time than the volume
of its input and output data implies - is, that we perform a lot of
intermediary calculations, later discarded. We  develop an algorithm
performing every calculation with a given item of input data just
once (or a constant number of times). First of all we notice that
Eq.(\ref{th2eq3}) can be rewritten as

\bea\label{interl1} \begin{cases}
{(m_{1L}^2+n_{1L}^2)}^{1/4} = {(m_{1R}^2+n_{1R}^2)}^{1/4} = \g_1\sqrt[4]{q_1}\\
{(m_{2L}^2+n_{2L}^2)}^{1/4} = {(m_{2R}^2+n_{2R}^2)}^{1/4} =\g_2\sqrt[4]{q_2}\\
m_{1L} - m_{1R} = -m_{2L} + m_{2R}\\
n_{1L} - n_{1R} = -n_{2L} + n_{2R}\\
\end{cases}\eea
where $q_1, q_2$ are two different class indexes and $\g_1, \g_2$ -
the corresponding weights.

\paragraph{Definition.} For any two decompositions of a number
$\g_1^4q$ into a sum of two squares (see (\ref{interl1})) the value
$\d_m=m_{L} - m_{R}$ is called $m$-{\it deficiency}, $\d_n=n_{L} -
n_{R}$ is called $n$-{\it deficiency} and $\vec{\d}_{m,n} = (\d_m,
\d_n)$ - {\it deficiency point}.\\

We immediately see that for two interacting waves their deficiencies
must be equal: $\d_{1m}=m_{1L} - m_{1R} = -m_{2L} + m_{2R} =
\d_{2m}, \quad \d_{1n} = n_{1L} - n_{1R} = -n_{2L} + n_{2R} =
\d_{2n}$. For a given weight $\g$, every two decompositions of
$\g^4q$ into a sum of two squares yield, in general, four deficiency
points with $\d_m, \d_n \ge 0$. Consider unsigned decompositions
$m_L, m_R, n_L, n_R \ge > 0$. Assuming $m_L \ge m_R, n_L \le n_R$
the four points are $(m_L+m_R, n_L+n_R), \quad (m_L+m_R, -n_L+n_R),
\quad (m_L-m_R, n_L-n_R), \quad (m_L-m_R, -n_L+n_R)$ and four
(symmetrical) points in each of the other three quadrants of the
$(m, n)$ plane.

\paragraph{Definition.} The set of all deficiency points of a
class for a given weight, $\Delta^\g_q$, is called its {\it
$\g$-deficiency set}. The set of all deficiency points of a class,
$\Delta_q$, is called its {\it deficiency set}.\\

The objects defined above play the main role in our algorithm, so we
compute as an illustrative example the for the number 50. The number
$50$ has three decompositions into sum of two squares, namely,
$50=1^2+7^2=5^2+5^2=7^2+1^2$, and nonnegative deficiency points of
decomposition pairs are $(5,5; 7,1),\quad (5,5; 1,7),\quad (1,7;
7,1).\quad $  They constitute a subset of the deficiency set
$\Delta^1_{50}$, namely, the 12 points with $m \ge 0, n \ge 0$. In
each of  three other quadrants of the $(m,n,)$ plane there lie 36
more points of this set, symmetrical to the ones shown with respect
to the coordinate axes.\\

The crucial idea behind the algorithm of this paper is very
simple and follows immediately from the exposition above:\\

{\bf Sys.(\ref{interl1}) has a solution with vectors belonging to
the two different classes  $Cl_{q_1}, Cl_{q_2}$ if and only if their
deficiency sets have a non-void intersection, $\Delta_{q_1} \cap
\Delta_{q_2} \neq \emptyset$, i.e. some elements belong to both classes.}\\.

\section{ALGORITHM DESCRIPTION}

Calculation of relevant class indexes $q$ by a sieve-like procedure,
admissible weights $\g$ and decomposition of $\g^4q$ into sum of two
squares have all been treated in full detail in \cite{KarKar}. One
new feature we introduced here is, that immediately after generating
the array of class bases $q$ we purge away those which, whatever the
admissible weight $\g$, do not have a decomposition into a a sum of
two squares $\g^4q = m^2+n^2$ with both $m \le D, \quad n \le D$.
For the problem considered in \cite{KarKar} this would be
superfluous because virtually all these classes have been anyhow
filtered away according to another criterium ($\mathcal{M}(q)=1,
\quad Dec(q) \le 4$) which does not apply here. In this way we
exclude 100562 classes from the 384145 which the sieving procedure returns.\\

Evidently for any deficiency point $\vec{\d}_{m,n}$ inequalities
$|\d_m| \le 2D, \quad |\d_n| \le 2D$ hold. And if deficiency sets of
two classes have a non-void intersection, they also have an
intersection over points with non-negative $|\d_m|, |\d_n|$. So we
start with declaring a two-dimensional array $arDeficiency(0..2D,
0..2D)$ of type byte which serves storing and processing deficiency
sets of the classes. The array is initialized with all zeroes.

\subsection{The Five-Pass Procedure}

\subsubsection{Pass 1: Marking deficiency points}
In the first pass for every class $q$ in the main domain $D$ we
generate its deficiency set $D_q$. Notice that after generating
deficiency set of the class for each weight $\g$ and uniting them we
must check for doubles and eventually get rid of them. Next, for
every deficiency point $(\d_m, \d_n)$ of the class we increment the
value of the corresponding element of the array by 1, except
elements with value 255 whose values are not changed.

\subsubsection{Pass 2: Discarding non-interacting classes}
In the second pass we generate deficiency sets once more and for
every point of the deficiency set of a class check the values of the
corresponding point of $arDeficiency$. If all these values are equal
to $1$, no waves of the class participate in resonant interactions
and the class is discarded from further considerations.\\

For the problem considered this pass excludes just a few ($313$)
classes, so the time gain is very modest. However, we include this
step into the presentation for two reasons. First of all, it {\it
had} to be done as no possibility of reducing the number of classes
considered as much as possible and as soon as possible (before the
most time-consuming steps) may be neglected. Second, though giving
not much gain for solution of the problem at hand, this elimination
techniques may play a major role in further applications of our
algorithm.

\subsubsection{Pass 3: Linking interaction points to interacting vectors}
In the third pass we generate a more detailed deficiency set for
each class, i.e. for all classes not discarded in the previous pass:
for every deficiency point $\vec{\d}_{m,n}$ we store $q, \g, m_L,
n_L, m_R, n_R$. We do {\bf not} discard duplicates as we did in the
previous two passes. Then we revisit the corresponding points of
$arDeficiency$ and to each point whose value is larger than $1$ link
the structure $(q, \g, m_L, n_L, m_R, n_R)$ described above.

\subsubsection{Pass 4: Gathering interaction points}
In the fourth pass we go through the array $arDeficiency$ once more
and store every point with value greater than one in an array
$arDeficiencySol(1..2D, 0..1)$. We also relink structures linked to
deficiency points to corresponding points of the new array.

\subsubsection{Pass 5: Extracting solutions}
The four passes above leave us with an array of points
$\vec{\d}_{m,n}$ and to each of these points a list of structures
$(q^i, \g^i, m^i_L, n^i_L, m^i_R, n^i_R)$ is linked (no less than
two different $q^i$). In general, a linked list is here most
appropriate. Every combination of two structures linked to the same
point and having different $q^i$ yields a solution of
Sys.(\ref{prosetdef2}). From every solution found, we obtain four
solutions changing signs of $m^i, n^i$ in the general case, i.e.
for $m^i, n^i$ nonzero.\\

Notice that theoretically we could skip Pass 4 and extract solutions
directly from the array $arDeficiency$. However, this is not
reasonable for implementation reasons, and Pass 4 is not very
time-consuming.

\subsubsection{Implementation remarks.}

Implementing the algorithm described above, we took a few
language-specific shortcuts that will be briefly described here.\\

Passes 1 and 2 have been implemented one-to-one as described above.
However, manipulating linked lists in VBA involves considerable
overhead and for the problem considered in this paper we do not need
the complete functionality of linked lists, i.e. inserting
into/deleting from intermediate positions of the list. Our main data
structure for Pass 3-5 is a simple two-dimensional array
arSolHalves($1..N_{MNdef}, 0..7$) and in a single line of this array
we store:
\begin{itemize}
\item{}the class base $q$;
\item{}the coordinates of deficiency point $d_m, d_n$;
\item{}the coordinates of two wave vectors belonging to this
deficiency point;
\item{}the index in the array arSolHalves of the next line belonging
to the same deficiency point.
\end{itemize}
which is demonstrated in Fig.\ref{f:ArrListA}  below.
\begin{figure}[h]
\centerline{\psfig{file=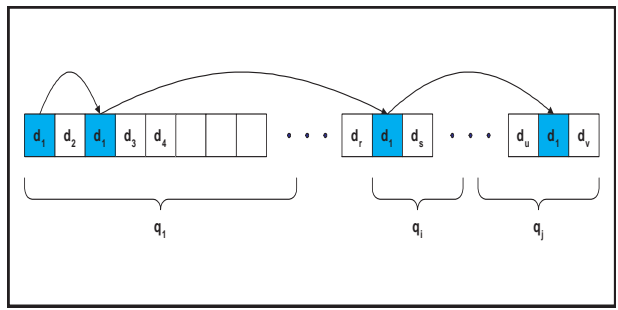,width=8cm,height=4cm}}
\vspace*{8pt} \caption{\bf Array simulation of deficiency point
lists: the overall array/list structure \label{f:ArrListA}}
\end{figure}
 Here, $N_{MNdef}$ is the number of $m,n$-vectors linked to all
deficiency points to which vectors belonging to two or more classes
belong ($6692832$ for $D=1000$). We generate the deficiency set of
each class and fill all members of this line of the array except the
last one in the process, deficiency point by deficiency point. The
last member is filled
later and in the following way.\\

\begin{figure}[h]
\centerline{\psfig{file=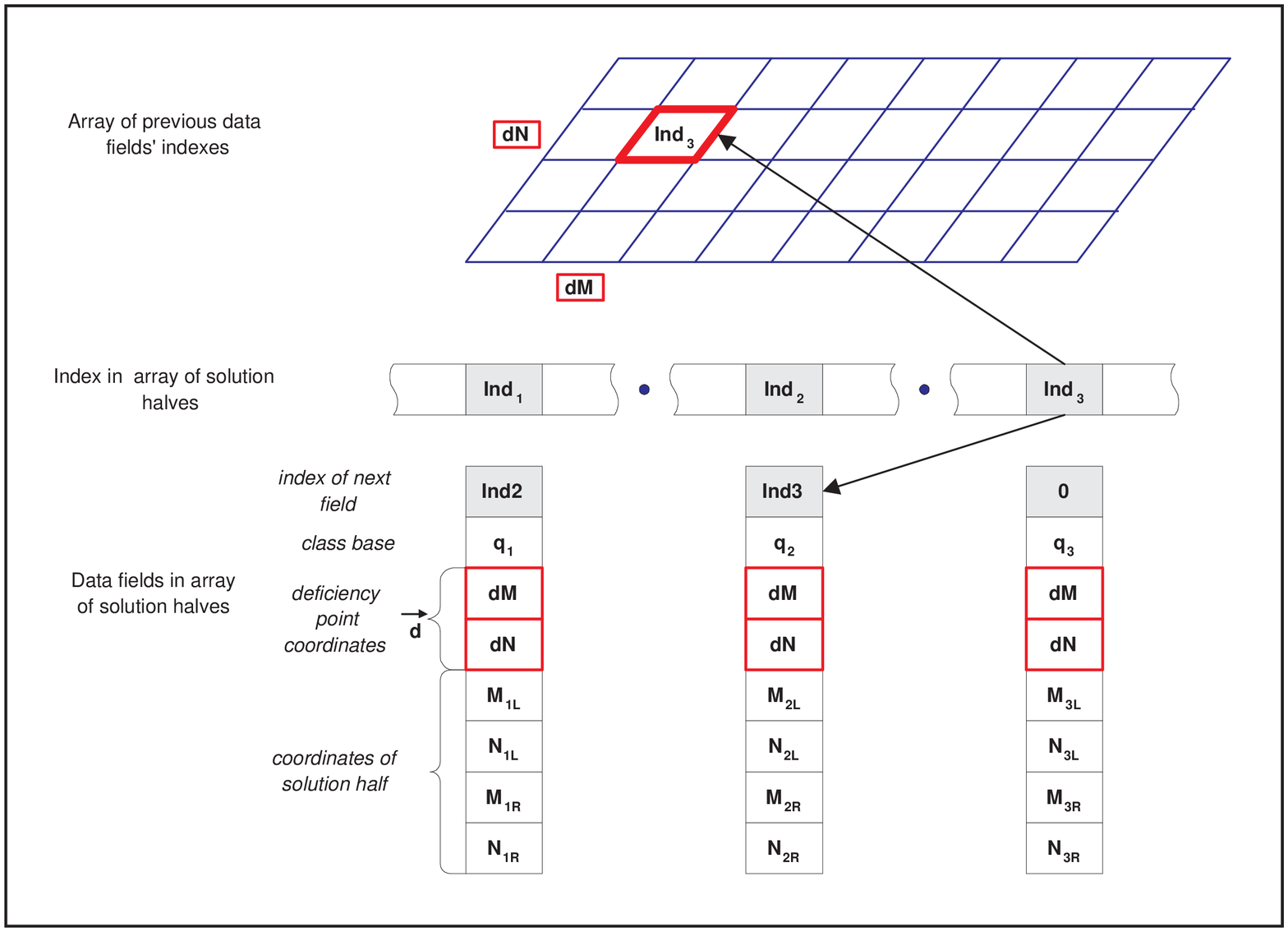, width=14cm,height=9cm}}
\vspace*{8pt} \caption{\bf Array simulation of deficiency point
lists: data fields in detail.\label{f:ArrListB}}
\end{figure}

For this pass we also declare an auxiliary array
arDeficiencesPrev(1..$2D$, 1..$2D$) initialized with zeroes. Having
added a new line $\{q, d_m, d_n, m_{1L}, n_{1L}, m_{1R}, n_{1R},
0\}$ to arSolHalves, we look up the value $ind_{d_m, d_n}$of
arDeficiencesPrev($d_m, d_n$). If it is zero (this deficiency point
being visited the first time) we just assign this point the value of
the index of the new line in the array arSolHalves. Otherwise we
first assign arSolHalves($ind_{d_m, d_n}$, 7) the value of the
current line's index in arSolHalves, then write this number to
arDeficiencesPrev($d_m, d_n$) (see Fig.\ref{f:ArrListB}).\\

A numerical example for this procedure is given in Table 1. In this
way, the array index of the next "list" member is stored with the
previous one, except evidently the last one, where the corresponding
field stays zero.

\begin{tabular}{|c|c|c|c|c|c|c|c|c|}
\hline 
    $Index$ & $q$ & $d_m$ & $d_n$ & $m_L$ & $n_L$ & $m_R$ & $n_R$ & $NextIndex$\\
\hline
 1  &  1  &  1  &  1  &  0  &  1  &  1  &  0  &  117 \\ \hline
 117  &  1  &  1  &  1  & -119  &  120  &  120  & -119  &  1241 \\ \hline
 1241  &  4  &  1  &  1  & -1  &  2  &  2  & -1  &  2921 \\ \hline
 2921  &  8  &  1  &  1  & -2  &  3  &  3  & -2  &  4958 \\ \hline
 4958  &  12  &  1  &  1  & -3  &  4  &  4  & -3  &  8107 \\ \hline
 8107  &  19  &  1  &  1  & -4  &  5  &  5  & -4  &  10304 \\ \hline
 ...  &  ...  &  ...  &  ...  &  ...  &  ...  &  ...  &  ...  &  ... \\ \hline
 6692782  &  273559  &  1  &  1  & -995  &  996  &  996  & -995  &  6692802 \\ \hline
 6692802  &  273567  &  1  &  1  & -996  &  997  &  997  & -996  &  6692816 \\ \hline
 6692816  &  273575  &  1  &  1  & -997  &  998  &  998  & -997  &  6692828 \\ \hline
 6692828  &  273580  &  1  &  1  & -998  &  999  &  999  & -998  &  6692832 \\ \hline
 6692832  &  273583  &  1  &  1  & -999  &  1000  &  1000  & -999  &  0 \\ \hline
\end{tabular}\\

{\small Table 1. A few lines of the table containing solution halves
for the deficiency point $\vec{d}=(1,1)$} (beginning and end of the sequence).\\

\subsection{Computational Complexity}

Consider computational complexity of these steps.

\subsubsection{Pass 1}
For a single class index $q$ and weight $\g$, generating deficiency
points in the first step consumes $\le O(\log^2(\g^4q))$ operations
because every number $X$ has no more than $O(\log X)$ decompositions
into two squares which we combine pairwise to find deficiency
points. Decompositions themselves can be found in $O(\log(\g^4q))$
time\cite{basil}. There are $(D/q)^{1/4}$ admissible weights to
class index $q$, so the overall complexity for a class can be
estimated from above as $\log^2DD^{1/4}$. Merging deficiency points
into $\Delta_q$ can be done in $O(X \log X)$ time for number of
points X, i.e. no more than
$O(\log^2DD^{1/4}\log(\log^2DD^{1/4})) = O(\log^3DD^{1/4})$ \\

Taking a rough upper estimate for the number of classes $O(D^2)$, we
obtain an estimate $O(\log^3DD^{2.25})$. Incrementing the points of
$arDeficiency$ is linear on the point number of the set $\Delta_q$
and need not be considered for computational complexity separately.

\subsubsection{Pass 2}
The same complexity estimate holds for the second pass. Notice that,
having enough memory, or using partial data loading/unloading
similar to that used in \cite{KarFolk}, we could preserve deficiency
sets calculated on the first pass and not recalculate them here.
However, this would not significally improve the overall
computational complexity of the algorithm.\\

We can not give an {\it a priori} estimate for the number of classes
discarded at the second pass, so we ignore it and hold the initial
rough upper estimate $O(D^2)$ for the number of classes in our
further considerations.

\subsubsection{Pass 3}
In the third pass, to every point $\d_{m,n}$ (no more than
$O(\log^2DD^{1/4})$ of them) we link the values $(q^i, \g^i, m^i_L,
n^i_L, m^i_R, n^i_R)$ for which this point has been struck. This, as
well as linking to the points of $arDeficiency$ is, clearly, linear
on the number of points and does not raise the computational
complexity.

\subsubsection{Pass 4}
Complexity of the fourth pass can be estimated as follows. Suppose
the worst case, i.e. no classes are discarded at step 2 and every
deficiency point is a solution point, i.e. for every $\vec{\d}_{m,n}
= (\d_m, \d_n)$ no less than two classes have deficiency points with
the same $d_m, d_n$. Then we must make no more than
$O(\log^2DD^{2.25})$ entries into the new array $arDeficiencySol$.
We must relink no more than the mean of $O(\log^2D)$ structures per
point, which gives an upper estimate of $O(\log^4DD^{2.25})$ time
for the pass. However, remember that the estimate for the deficiency
point number has been made on the assumption that all $(m^i_L,
n^i_L, m^i_R, n^i_R)$ generate distinct deficiency points. In simple
words, for every point linked to $X>1$ structures we obtain $X-1$
less solution points. Now elementary consideration allow us to
improve the estimate to $O(\log^2DD^{2.25})$ time.

\subsubsection{Pass 5}
We did not manage to obtain a reasonable estimate for the
computational complexity of the fifth step. For the worst case of
all structures grouped at a single  point, the estimate is
$O(\log^4DD^{4.5})$ - but this is not realistic. If the number of
solution points is $O(\log^2DD^{2.25})$ and the number of linked
deficiencies is bounded by some number $c$, then we can make an
estimate $O(c^2\log^2DD^{2.25})$. This, however, is also not quite
the case as our numerical simulations show). However, this last step
deals with solution extraction and extracts them in linear time per
solution. {\it Any} algorithm solving the problem has to extract
solutions, so we can be sure that our step 5 is optimal - even
without any estimate of its computational complexity. Summing up, we
obtain the overall upper estimate of computational complexity
$O(\log^3DD^{2.25})$ reached at steps $1$ and $2$ plus the time
needed for solution extraction.

\section{DISCUSSION}
\begin{figure}[h]
\centerline{\psfig{file=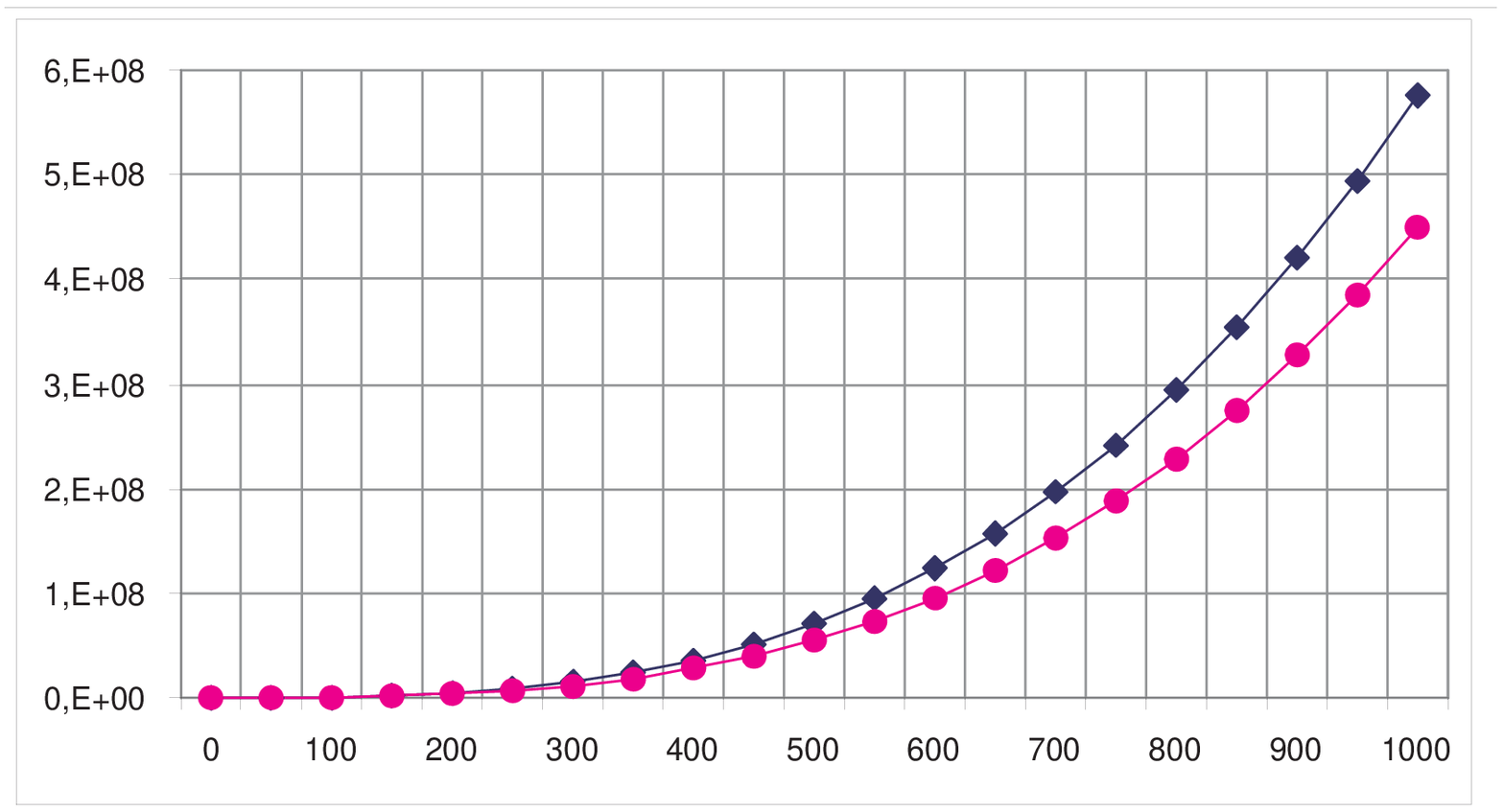, width=10cm,height=5cm}}
\vspace*{8pt} \caption{\bf Number of solutions in partial domains
$m_i, n_i \le D$ (curve marked diamonds) and $m_i^2+n_i^2 \le D^2$
(curve marked circles), $D=50, 100,... 1000$. \label{f:SolSqSir}}
\end{figure}

Our algorithm has been implemented in the VBA programming language;
computation time (without disk output of solutions found) on a
low-end PC (800 MHz Pentium III, 512 MB RAM) is about 10 minutes.
Some overall numerical data is given in the two Figures below. The
number of solutions for the 2-class-case depending on the partial
domain is shown in the Fig.\ref{f:SolSqSir}. Both curves are almost
ideal cubic lines. Very probably they {\it are}
cubic lines asymptotically - the question is presently under study.\\

Partial domains chosen in Fig.\ref{f:SolSqSir} are of two types:
squares $m_i, n_i \le D$, just for simplicity of computations, and
circles $m_i^2+n_i^2 \le D^2$, more reasonable choice from physical
point of view (in each circle all the wave lengths are $\le D$). The
curves in the Fig.\ref{f:SolSqSir} are very close to each other in
the domain $D \le 500$ though number of integer nodes in a
corresponding square is $D^2$ and in a circle with radius $D$ there
are only $\pi D^2/4$ integer nodes. This indicates a very
interesting physical phenomenon: most part of the solutions is
constructed with the
wave vectors parallel and close to either axe $X$ or axe $Y$. \\

On the other hand, the number of solutions in rings $ (D-50)^2 <
m_i^2+n_i^2 \le D^2$ (corresponds to the wavelengths between
 $D-50$ and $D$) grows nearly
perfectly linearly. Of course the number of solutions in a circle is
{\it not} equal to the sum of solutions in its rings: a solution
lies in some ring if and only if all its four vectors lie in that
same ring. That is, studying solutions in the rings only, one
excludes automatically a lot of solutions containing vectors with
substantially different wave lengths simultaneously, for example,
with wave vectors from the rings $D-50$ and $D+100$. This "cut" sets
of solutions can be of use for interpreting of the results of
laboratory experiments performed for investigation of waves with
some given wave lengths (or frequencies) only.

\begin{figure}[h]
\centerline{\psfig{file=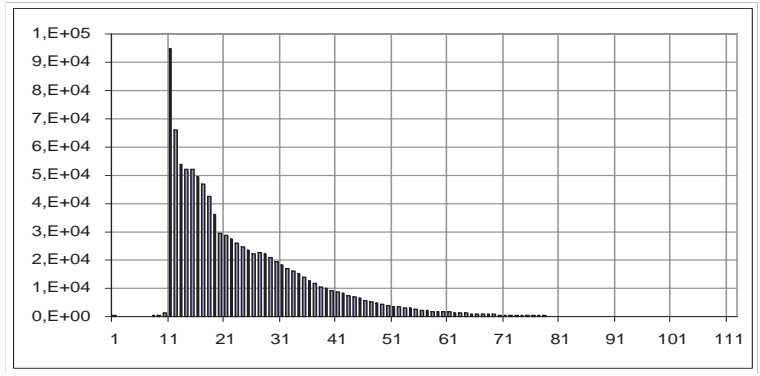, width=10cm,height=5cm}}
\vspace*{8pt} \caption{\bf The multiplicities histogram.
\label{f:VecMul222}}
\end{figure}

Another important characteristic of the structure of the solution
set is multiplicity of a vector which describes how many times a
given vector is a part of some solution. The multiplicity histogram
is shown in Fig.\ref{f:VecMul222}. On the axis $X$ the multiplicity
of a vector is shown and on the axis $Y$ the number of vectors with
a given multiplicity. The histogram of multiplicities is presented
in Fig. \ref{f:VecMul222}, it has been cut off - multiplicities go
very high, indeed the vector (1000,1000) takes part in 11075
solutions.\\

Similar histograms computed for different 1-class-cases show that
most part of the vectors, $70-90\%\%$ for different types of waves,
take part in one solution, e.g. they have multiplicity 1.  This
means that triads or quartets are, so to say, the "primary elements"
of a wave system and we can explain its most important energetic
characteristics in terms of these primary elements. The number of
vectors with larger multiplicities decreases exponentially when
multiplicity is growing. The very interesting fact in the
2-class-case is the existence of some initial interval of small
multiplicities, from 1 to 10, with very small number of
corresponding vectors. For instance, there exist only 7 vectors with
multiplicity 2. Beginning with multiplicity 11, the histogram is
similar to that in the 1-class-case.\\

This form of the histogram is quite unexpected and demonstrates once
more the specifics of the 2-class-case compared to the 1-class-case.
As one can see from the multiplicity diagram in Fig.
\ref{f:VecMul222}, the major part in 2-class-case is played by much
larger groups of waves with the number of elements being of order
40:  each solution consists of 4 vectors, groups contain at least
one vector with multiplicity 11 though some of them can take part in
the same solution. This sort of primary elements can be a
manifestation of a very interesting physical phenomenon which should
be investigated later:
 triads and quartets as primary elements demonstrate
periodic behavior and therefore the whole wave system can be
regarded as a quasi-periodic one. On the other hand, larger groups
of waves may have chaotic behavior and, being primary elements,
define quite different way of energy transfer through the whole wave
spectrum.

\end{document}